\documentclass[aps,prl,twocolumn,showpacs,superscriptaddress]{revtex4}
\usepackage{epsfig}
\usepackage{graphicx}
\usepackage{bm}
\begin{document}
\date{\today}

\title{Scheme for generating coherent state superpositions with realistic cross-Kerr nonlinearity}

\author{Bing He} 
\email{bhe98@earthlink.net}
\affiliation{Department of Physics and Astronomy, Hunter College of the City University of New York, 695 Park Avenue, 
New York, NY 10065, USA}
\author{Mustansar Nadeem}
\affiliation{Department of Physics, Quaid-i-Azam University, Islamabad 45320, Pakistan}
\author{J\'{a}nos A. Bergou} 
\affiliation{Department of Physics and Astronomy, Hunter College of the City University of New York, 695 Park Avenue, 
New York, NY 10065, USA}


\begin{abstract}
We present a simple scheme using two identical cross-phase modulation processes in decoherence environment to generate superpositions of two coherent states with the opposite phases, which are known as cat states. The scheme is shown to be robust against decoherence due to photon absorption losses and other errors, and the design of its experimental setup is 
also discussed.

\end{abstract}

\date{\today}
\maketitle

Schr\"odinger's famous cat paradox can be realized by optical coherent state superpositions in the form 
of $|CSS_{\pm}(\beta)\rangle=N_{\pm}~(|\beta\rangle \pm |-\beta\rangle)$ ($|\pm\beta\rangle$ is coherent state with the amplitude $|\beta|$ and $N_{\pm}=(2\pm 2~ exp~[-2|\beta|^2])^{-\frac{1}{2}}$). 
$|CSS_{+}\rangle$ ($|CSS_{-}\rangle$) is called an even (odd) cat state, since it is the superposition of even (odd) photon number states. Cat states and other coherent state superpositions have been proposed to implement various quantum information tasks such as linear-optics quantum computation \cite{ralph, lund} and quantum metrology \cite{ralph2, gilchrist, m-b}. Generation of these states is therefore under intensive research  recently (see \cite{g-v} for a comprehensive review).

One line of research in the field is to generate cat states through a cross-phase modulation (XPM) process in Kerr medium 
\cite{v, gerry, jeong}. Such an ideal process is described by the Hamiltonian $H=-\hbar\chi \hat{a}^{\dagger}\hat{a} \hat{b}^{\dagger}\hat{b}$, where $\chi$ is the nonlinear strength, and $\hat{a}$ and $\hat{b}$ two coupling optical modes. A simple approach of this kind is Gerry's scheme \cite{gerry}, where an input coherent state $|\alpha\rangle_1$ as the probe in Fig. 1 interacts with one of the single photon modes, $|0,1\rangle_{2,3}\equiv |0\rangle$ and  $|1,0\rangle_{2,3}\equiv |1\rangle$, as the signal through an XPM process, and the state of the coherent beam is post-selected to a cat state by the detection of the single photon mode $D_1$ or $D_2$ if the XPM phase $\theta$ could be $\pi$. Simple though the scheme is, realizing a large $\theta$ is still challenging with the current technology. Even with electromagnetically induced transparent (EIT) material \cite{eit}, the initially achieved phase shift $\theta$ at the single photon level is only in the order of $10^{-5}$ \cite{s-i}. Moreover, all Kerr nonlinear materials carry a complex third order susceptibility $\chi^{(3)}=Re \chi^{(3)}+iIm\chi^{(3)}$, necessitating the decays of the coupling optical modes caused by the imaginary part. Under the decoherence effect caused by such losses, the XPM processes of the density matrix $\rho$ of an involved system are quantum operations (QOs) described by the master equation
\begin{eqnarray}
\frac{d\rho}{dt}=i\hbar\chi\sum_{i,j}[\hat{a}^{\dagger}_i\hat{a}_i \hat{a}_j^{\dagger}\hat{a}_j, \rho]+\frac{\gamma}{2}\sum_{i}\{[\hat{a}_i\rho,\hat{a}_i^{\dagger}]+[\hat{a}_i,\rho\hat{a}_i^{\dagger}]\},~~
\label{1}
\end{eqnarray}
rather than the ideal unitary transformations. The nonlinear strength $\chi$ and the damping rate $\gamma$ of the coupling optical modes $\hat{a}_i$ are from the real and the imaginary parts of $\chi^{(3)}$, respectively. 

\begin{figure}
\includegraphics[width=84truemm]{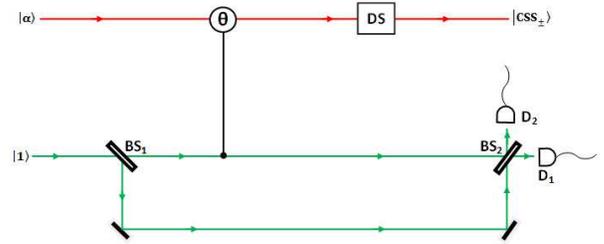}
\caption{(color online) The setup of single XPM scheme, where the input qubit and coherent state are in the state $2^{-\frac{1}{2}}(|0\rangle+|1\rangle)|\alpha\rangle$. An ideal XPM process inducing $\theta$ transforms the input to $2^{-\frac{1}{2}}|0\rangle |\alpha\rangle+2^{-\frac{1}{2}}|1\rangle|\alpha e^{i\theta}\rangle$. The displacement DS is necessary here for generating the coherent state superpositions with a weak cross-Kerr nonlinearity. The XPM process should be treated as a quantum operation if one considers photon absorption losses.} 
\end{figure}

To generate cat states with weak cross-Kerr nonlinearity, Jeong proposed applying the idea of compensating for a small $\theta$ with a large intensity of the input coherent beam in \cite{cc} and obtained a post-selected mixed state \cite{jeong}
\begin{eqnarray}
\rho_{\pm}(t)&\sim&|A\alpha\rangle\langle A\alpha|\pm C(t)|A\alpha\rangle\langle A\alpha e^{i\theta}|\pm C^{\ast}(t)|A\alpha e^{i\theta}\rangle\langle A\alpha| \nonumber\\
&+&|A\alpha e^{i\theta}\rangle\langle A\alpha e^{i\theta}|\nonumber\\
&=& \frac{1+|C(t)|}{2}|CG_{\pm}\rangle\langle CG_{\pm}|
+\frac{1-|C(t)|}{2}|CG_{\mp}\rangle\langle CG_{\mp}|~~~~
\label{2}
\end{eqnarray}
under the decoherence effect, where $|CG_{\pm}\rangle=|A \alpha\rangle\pm \{C^{\ast}(t)/|C(t)|\}|A \alpha e^{i\theta}\rangle$, $A=e^{-\frac{\gamma}{2}t}$, and the closed form of the complex coherence parameter $C(t)$ is given in \cite{l-n}. After one performs a displacement $D(x)$ such that $|A \alpha\rangle\rightarrow |\beta\rangle$ and $|A \alpha e^{i\theta}\rangle\rightarrow |-\beta\rangle$, the pure state components $|CG_{\pm}\rangle$ will be transformed to  $|CG'_{\pm}\rangle\sim|\beta\rangle\pm e^{iarg C^{\ast}} e^{i\phi_D}|-\beta\rangle$ ($\phi_D$ is the relative phase from the displacement). Small $\theta$ can be therefore compensated by a large $|\alpha|$ so that 
the amplitude $|\beta|$ of the realized $|CG'_{\pm}\rangle$ could be big enough. 

One problem in the scheme is the implementation of the displacement $D(x)$ on an intense coherent beam in the state of 
Eq. (\ref{2}). A displacement on a coherent state can be approximated by a beam splitter of extremely high transmissivity, which is fed by a very intense coherent beam  at the second port \cite{paris}. However, if one wants to displace the state by a very large $|x|$ as in \cite{jeong}, the intensity of the second beam would be beyond the reasonable value. 

The other problem is the difference of the generated $|CG'_{\pm}\rangle$ from an even or odd cat state $|CSS_{\pm}\rangle$ by a relative phase $\phi=arg C^{\ast}+\phi_D$ arising from the decoherence and the displacement. Directly changing the relative phase $\phi$ requires some type of nonlinear interaction \cite{g-v}. There are two ways to convert $|CG'_{\pm}\rangle$ to $|CSS_{\pm}\rangle$ with only linear optics: one is to prepare the state of the input single photon qubit as $|0\rangle+e^{-i\phi}|1\rangle$ to cancel that of the coherent states; the other is to have two such states $|CG'_{1,+}\rangle=|\beta\rangle+ e^{i\phi_1}|-\beta\rangle$ and $|CG'_{2,+}\rangle=|\beta\rangle+ e^{i\phi_2}|-\beta\rangle$ satisfying $\phi_1+\phi_2=\pi$, and transform them together by a beam splitter to $|\sqrt{2}\beta\rangle+ |-\sqrt{2}\beta\rangle$ \cite{am}. By these methods the perfect match of the qubit and the coherent state relative phases (or two relative phases of the coherent states) is necessary, so the scheme is sensitive to these independently created phases in generating $|CSS_{\pm}\rangle$. Moreover, $|1\rangle$ component of the single photon picks up an extra phase $\phi_E$ and a decay factor due to the different propagation velocity from that of $|0\rangle$ component and its loss in nonlinear medium, adding more consideration to the experimental realization of the scheme. 

\begin{figure}
\includegraphics[width=84truemm]{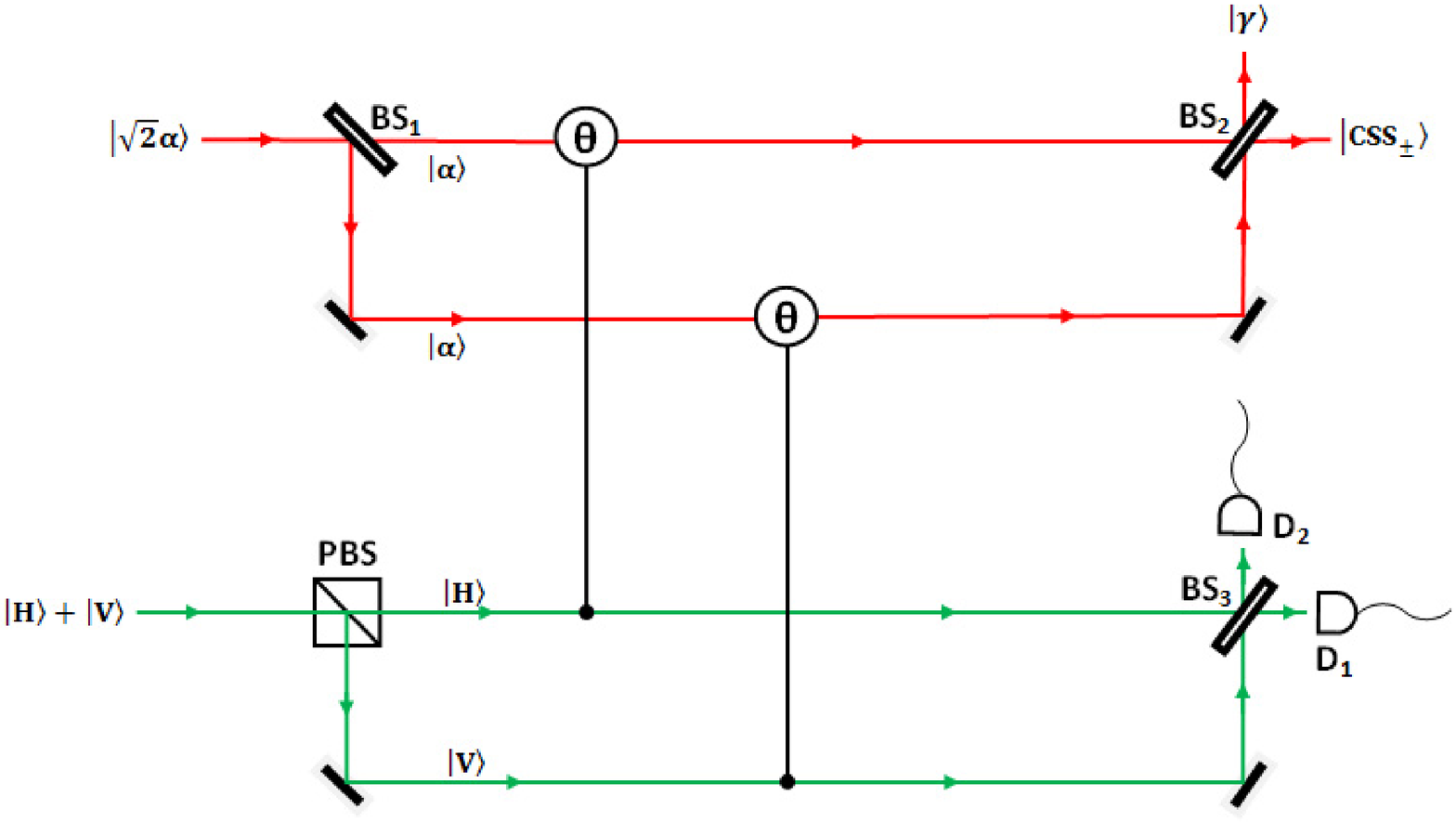}
\caption{(color online) Improved setup for generating even and odd cat states. Under the decoherence from photon absorption losses, two ideal unitary transformations of the XPM processes should be substituted by the effective actions of two quantum operations, which map the input to a mixture of even and odd cat states conditioned on the detection of $D_1$ or $D_2$ mode.} 
\end{figure}

Here we present a scheme of double XPM outlined in Fig. 2 to overcome the above-mentioned shortcomings in single XPM scheme (a similar scheme without considering the photon absorption losses in XPM processes is given in \cite{k-p}).
By {\it double XPM} we mean two identical XPM processes inducing the same phase $\theta$. We choose the single photon state as the superposition of two polarizations, $2^{-\frac{1}{2}}(|H\rangle+|V\rangle)$ ($H$ and $V$ are horizontal and vertical polarization, respectively), but the effect will be the same if we use the single photon state of Fig. 1. Separated by a 50/50 beam splitter $BS_1$ and a polarization beam-spltter (PBS), the coherent beam and the single photon as the whole system will be in the following input state
\begin{eqnarray}
|\Psi\rangle_{in}=\frac{1}{\sqrt{2}}\left(|H\rangle_3+|V\rangle_4\right)|\alpha\rangle_1|\alpha\rangle_2.
\end{eqnarray}

To study its evolution determined by Eq. (\ref{1}), we should consider four modes corresponding to two coherent beams and $H$/$V$ polarization of the single photon, and the first summation $\sum_{i,j}$ in Eq. (\ref{1}) will be over the modes $(1,3)$ and $(2,4)$. We here use the following operators
\begin{eqnarray}
{\cal K}_{ij}\rho&=&i\hbar\chi[\hat{a}^{\dagger}_i\hat{a}_i \hat{a}_j^{\dagger}\hat{a}_j, \rho],\nonumber\\
{\cal J}_i \rho&=&\frac{\gamma}{2}[\hat{a}_i\rho,\hat{a}_i^{\dagger}],
~~~~~~{\cal L}_i \rho=\frac{\gamma}{2}[\hat{a}_i,\rho\hat{a}_i^{\dagger}],
\end{eqnarray}
for simplicity.
By dividing the interaction time into infinitely many small periods, we express the QOs on the input, $\rho(t_0)=|\Psi\rangle_{in}\langle\Psi|$, from $t_0$ to $t$ as follows:
\begin{eqnarray}
&&\rho(t)=\lim_{N\rightarrow \infty}\prod_{k=1}^{N-1}\underbrace{(I+\sum_i({\cal J}_i+{\cal L}_i )\Delta t)}\limits_{{\cal D}(t_k)}\underbrace{(I+\sum_{i,j}{\cal K}_{i,j}\Delta t)}\limits_{{\cal U}(t_k)}\rho(t_0),\nonumber\\
\end{eqnarray}
with $\Delta t=(t-t_0)/N$ and $t_k=t_0+k\Delta t$.
The first small step of operation ${\cal D}(t_1){\cal U}(t_1)$ maps $\rho(t_0)$ to (the mode indexes are neglected)
\begin{widetext}
\begin{eqnarray}
\rho(t_1)&=& \frac{1}{2} {\cal D}(t_1){\cal U}(t_1)\{(|H\rangle+|V\rangle)(\langle H|+ \langle V|)\otimes |\alpha\rangle\langle\alpha|\otimes
|\alpha\rangle\langle\alpha|\}\nonumber\\
&=&\frac{1}{2}{\cal D}(t_1)\left(|H\rangle|\alpha e^{i\chi \Delta t}\rangle|\alpha\rangle+|V\rangle |\alpha\rangle |\alpha e^{i\chi \Delta t}\rangle\right)
\left(\langle H| \langle\alpha e^{i\chi \Delta t}
|\langle\alpha|+\langle V|\langle \alpha|\langle\alpha e^{i\chi \Delta t}|\right)\nonumber\\
&\sim&|H\rangle\langle H|\otimes|\alpha e^{(-\frac{\gamma}{2}+i\chi) \Delta t},\alpha e^{-\frac{\gamma}{2}\Delta t}\rangle \langle\alpha e^{(-\frac{\gamma}{2}+i\chi) \Delta t},\alpha e^{-\frac{\gamma}{2}\Delta t}|+|V\rangle\langle V|\otimes|\alpha e^{-\frac{\gamma}{2}\Delta t},\alpha e^{(-\frac{\gamma}{2}+i\chi) \Delta t}\rangle \langle\alpha e^{-\frac{\gamma}{2}\Delta t},\alpha e^{(-\frac{\gamma}{2}+i\chi) \Delta t}|\nonumber\\
&+&C_1|H\rangle\langle V|\otimes|\alpha e^{(-\frac{\gamma}{2}+i\chi) \Delta t},\alpha e^{-\frac{\gamma}{2}\Delta t}\rangle \langle\alpha e^{-\frac{\gamma}{2}\Delta t},\alpha e^{(-\frac{\gamma}{2}+i\chi) \Delta t}|\nonumber\\
&+&C_1|V\rangle\langle H|\otimes|\alpha e^{-\frac{\gamma}{2}\Delta t},\alpha e^{(-\frac{\gamma}{2}+i\chi) \Delta t}\rangle \langle \alpha e^{(-\frac{\gamma}{2}+i\chi) \Delta t},\alpha e^{-\frac{\gamma}{2}\Delta t}|,
\label{6}
\end{eqnarray}
\end{widetext}
where $C_1=exp\{-(1-e^{-\gamma\Delta t})|\alpha e^{i\chi\Delta t}-\alpha|^2\}$. 
Two identical unitary operations between the modes $(1,3)$ and $(2,4)$ in ${\cal U}(t_1)$ create a symmetric form of the coherent states on the second line of Eq. (\ref{6}), and then the phases gained from ${\cal D}(t_1)$ for the off-diagonal qubit terms,  $|H\rangle\langle V|$ and $|V\rangle\langle H|$, can be canceled to obtain a real number $C_1$. The approximation on the third line of Eq. (\ref{6}) is the negligence of a common decay factor for all four qubit terms from the symmetric XPM processes. The $k$-th step operation ${\cal D}(t_k){\cal U}(t_k)$ contributes a similar coefficient $C_k$.
The QOs of the XPM processes therefore map $\rho(t_0)$ to $\rho(t)$, which can be decomposed to 
 \begin{eqnarray}
\rho(t)\sim\frac{1+C(t)}{2}|CS_{+}\rangle\langle CS_{+}|
+\frac{1-C(t)}{2}|CS_{-}\rangle\langle CS_{-}|,~
\label{8}
\end{eqnarray}
where 
\begin{eqnarray}
|CS_{\pm}\rangle&=&|H\rangle|e^{(-\frac{\gamma}{2}+i\chi)t}\alpha \rangle|e^{-\frac{\gamma}{2}t}\alpha \rangle\nonumber\\
&\pm& |V\rangle 
|e^{-\frac{\gamma}{2}t}\alpha\rangle|e^{(-\frac{\gamma}{2}+i\chi)t}\alpha \rangle.
\label{9}
\end{eqnarray}
The coherence parameter 
\begin{eqnarray}
&&C(t)=\lim_{N\rightarrow \infty} C_1C_2\cdots C_{N-1}\nonumber\\
&=&exp\{-2|\alpha|^2(\frac{\chi^2}{\gamma^2+\chi^2}-e^{-\gamma t}+\frac{\gamma^2}{\gamma^2+\chi^2}e^{-\gamma t}\cos\chi 
t\nonumber \\
&&-\frac{\gamma\chi}{\gamma^2+\chi^2}e^{-\gamma t}\sin\chi t)\}
\label{10}
\end{eqnarray}
is shown in Fig. 3. Two 50/50 beam splitters $BS_2$ and $BS_3$ are applied to transform the coherent and the single photon modes in Eq. (\ref{9}) to the proper forms. If one detects the single photon mode $D_1$ or $D_2$ in Fig. 2, the generated state will be the mixture of an even and an odd cat state $|CSS_{\pm}(\beta)\rangle$, with its size given 
as 
\begin{eqnarray}
|\beta|^2=2e^{-\gamma t}\sin^2\frac{\chi t}{2}|\alpha|^2,
\label{11}
\end{eqnarray}
and fidelity as $F=(1+C(t))/2$. Another interesting thing is that a pure coherent state $|\gamma\rangle=|\frac{e^{-\frac{\gamma}{2}t+i\chi t}+e^{-\frac{\gamma}{2}t}}{\sqrt{2}}\alpha\rangle$ outputs from 
the other port in Fig. 2, as long as the two XPM processes are symmetric.

\begin{figure}
\includegraphics[width=80truemm]{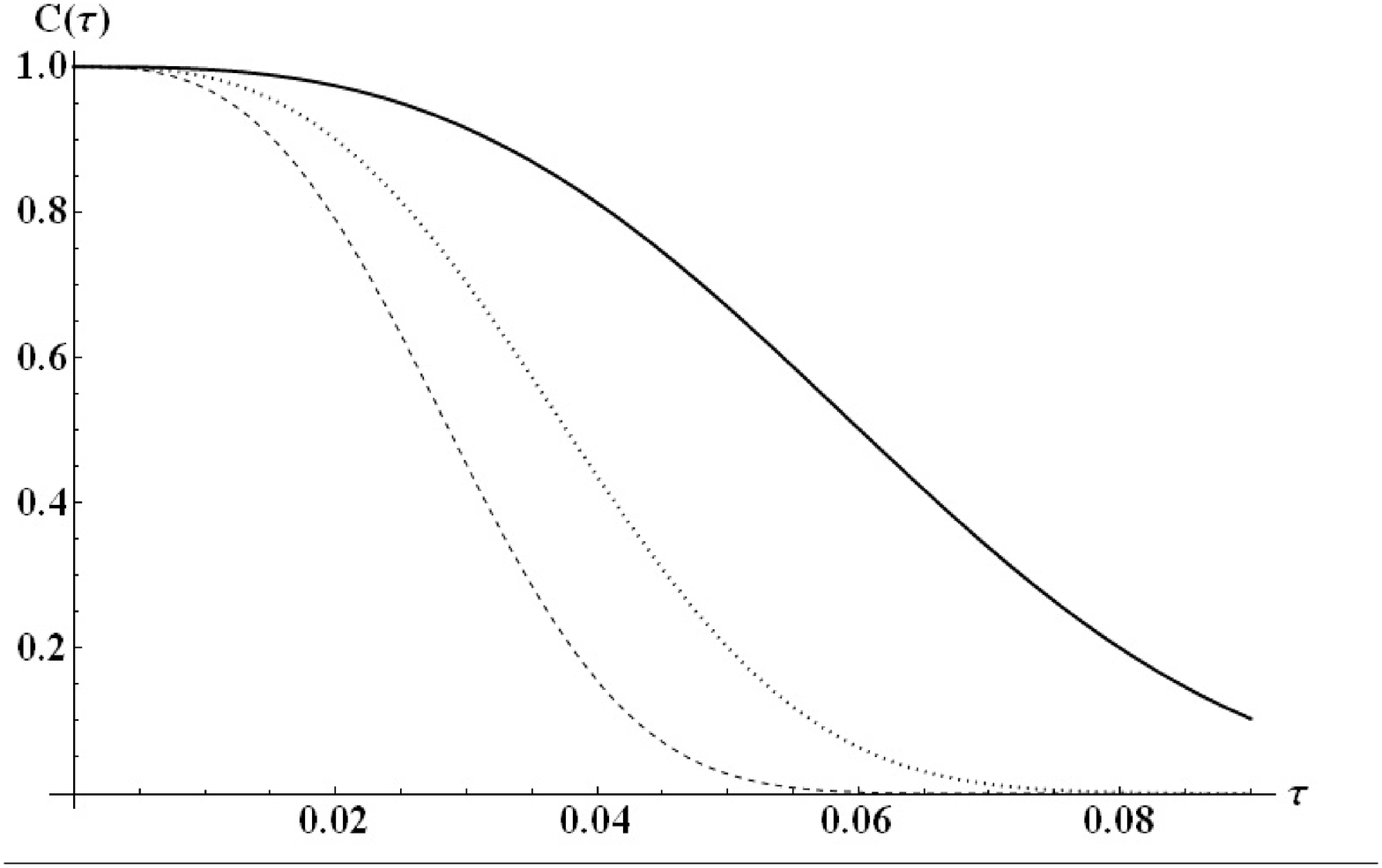}
\caption{The coherent parameter $C(\tau)$ vs dimensionless time $\tau$ for $|\alpha|=200$. The solid, dotted and dashed 
line represent the cases of $\Gamma=0.5$, $1$ and $1.5$, respectively.} 
\end{figure}

There are two primary advantages in the scheme: (1) the precise displacement of the strong coherent beam in Fig. 1 is replaced by two XPM processes, which could be easier to implement; (2) the coherence parameter $C(t)$ is real instead of 
the complex one in \cite{jeong}, so it is unnecessary to apply other procedures to convert the output to an even or odd cat state. The essential point is that we let two groups of the input optical modes, $\{|\alpha\rangle_1, |H\rangle\}$ and $\{|\alpha\rangle_2, |V\rangle\}$, undergo the same physical process in nonlinear Kerr medium, so the photon absorption decoherence on both of the groups and the phases gained by two coherent states and two single photon components should be identical. To achieve the target, we should have a well stabilized setup to process the inputs.  

In setting up the circuit in Fig. 2, one could meet with the asymmetry of two XPM processes. As the result, the symmetric pure state components in Eq. (\ref{9}) will become
\begin{eqnarray}
|CS'_{\pm}\rangle&=&|H\rangle|e^{(-\frac{\gamma}{2}+i\chi)t_1}\alpha \rangle|e^{-\frac{\gamma}{2}t_2}\alpha \rangle\nonumber\\
&\pm& e^{i\phi'_E}|V\rangle 
|e^{-\frac{\gamma}{2}t_1}\alpha\rangle|e^{(-\frac{\gamma}{2}+i\chi)t_2}\alpha \rangle,
\label{9-b}
\end{eqnarray}
with the different interaction times $t_1$ and $t_2$ for the two groups of optical modes, which also give rise to a relative phase $\phi'_E$ in the above equation. The resilience of a double XPM scheme to such asymmetry is estimated in \cite{k-p}:
for a deviation of $\theta_1=\chi t_1$ and $\theta_2=\chi t_2$ as large as $10\%$, the fidelity of the output is still larger than $0.95$. Here we propose a direct way to test how good the symmetry in circuit is. We just prepare a coherent state $|\gamma\rangle$, which is identical with that output from the other port in Fig. 2 in case of symmetry, according to the target cat state size, and then compare it with the coherent state components output from the second port by means of a 50/50 beam splitter and a simple photodiode as in \cite{andersson}. The difference of two coherent states $|\gamma\rangle$ and $|\gamma'\rangle$ due to the asymmetry can be well identified with an efficiency of $1-exp~(-\frac{1}{2}|\gamma-\gamma'|^2)$. 

We now look at the design of the setup to generate a cat state with a fidelity $F=(1+C)/2$ and an amplitude $|\beta|$. 
From Eqs. (\ref{10}) and (\ref{11}) we obtain a relation
\begin{eqnarray}
|\beta|^2 G(\tau)&=&|\beta|^2\frac{1-\frac{\Gamma^2}{1+\Gamma^2}e^{\tau}-\frac{1}{1+\Gamma^2}\cos \Gamma \tau+\frac{\Gamma}{1+\Gamma^2}\sin\Gamma \tau}
{\sin^2\frac{\Gamma \tau}{2}}\nonumber\\
&=&\ln (2F-1),
\label{12}
\end{eqnarray}
where $\tau=\gamma t$ and $\Gamma=\chi/\gamma$.
For any value of $\Gamma$, $G(\tau)$ ranges from $-\infty$ to $0$, as demonstrated with the representative $\Gamma$ values in Fig. 4. From Eq. (\ref{12}) we can definitely find a dimensionless interaction time $\tau_{int}$ for coherent beam and single photon if we specify any fidelity $F$ and any size $|\beta|^2$ for the generated cat state. 
For a sufficiently high fidelity $F=1-x$ ($0<x\ll 0.1$), this dimensionless interaction time $\tau_{int}$ is simply determined by $G(\tau_{int})=-2x/|\beta|^2$, and should be very small if $|\beta|$ is also large enough. We thus draw a conclusion: the only way to create cat states of high fidelity and large size through XPM is coupling a sufficiently strong coherent beam to a single photon within a limited time in Kerr medium. This is valid to any cross-Kerr nonlinearity and to single XPM scheme \cite{jeong} as well.

\begin{figure}
\includegraphics[width=80truemm]{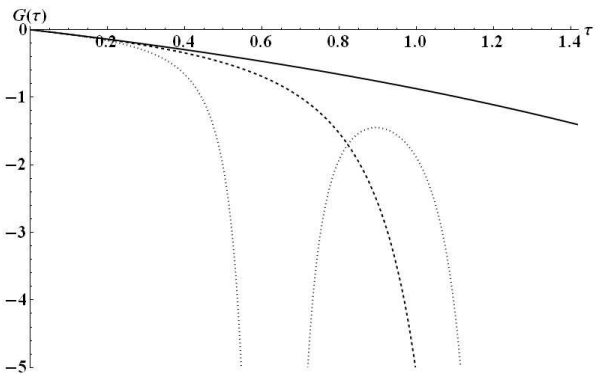}
\caption{Function $G(\tau)$ vs dimensionless time for three values of $\Gamma=0.01$, $5$ and $10$, which are represented 
by solid, dashed and dotted lines, respectively.} 
\end{figure}

For the example of the generated cat state with an amplitude $|\beta|=1.6$ and 
a fidelity $F=0.99$, we provide the following table 
of $\tau_{int}$ and input beam intensity $|\alpha|^2$:
\begin{center}
\begin{tabular}{|c|c|c|c|c|c|c|}\hline
$\Gamma$ & 0.01 & 1 & 25 & 50 & 100\\ \hline
$\tau_{int}$ & 0.0116846 & 0.011685 & 0.011652 & 0.011555& 0.011196 \\ \hline
$|\alpha|^2$& $1.2 \times 10^{12}$& $1.2 \times 10^{8}$&$2.0 \times 10^{5}$ & $5.1 \times 10^{4}$&$1.4 \times 10^{4}$ \\ \hline
\end{tabular}
\end{center}
The $\Gamma$ values range from that of normal silica core fiber to those achievable in EIT materials. The necessary input coherent beam intensity drops quickly with the increased $\Gamma$, and the $\tau_{int}$ values are very close because the $G(\tau)$ curves for the different $\Gamma$ values stick together near the origin as shown in Fig. 4. 

The data in the table also reflects the trade-off between the necessary nonlinear strength and the required coherent beam intensity similar to that in \cite{cc}. For a realistic system, however, the intensity of the coherent beam can not be boundlessly large because a very strong beam might also cause other effects. Good candidates for weak cross-Kerr nonlinearity without self-phase modulation effect are atomic systems working under EIT conditions. The $\Gamma$ parameters of EIT or double EIT systems are the ratio of the signal field detuning to the decay rate of the excited state \cite{p-k-h, m-s} 
(or to a quantity related to this decay rate for light-storage XPM approach \cite{chen-yu}). To create EIT condition in hot atoms, e.g., the probe beam should be weak enough (at the $\mu $W level) while the coupling beam is strong (at the $m$W level). By choosing a proper detuning, we could use a probe beam of the intensity $2|\alpha|^2$ consistent with such requirement in EIT systems to implement the scheme. The other issues, such as controlling the interaction time $t_{int}=\tau_{int}/\gamma$ in optical cavity or by other methods, still await further research.   

In summary, we have studied a double XPM scheme to generate cat states in decoherence environment. We show that this scheme is robust against photon absorption losses and other errors. The results obtained here are also applicable to the design 
of other quantum non-demolition detection (QND) setups based on XPM process.   

B. H. thanks Dr. W.-Z. Tang, Prof. I. A. Yu for discussions on experimental requirement for XPM process in EIT materials;
M. N. is sponsored by IRSIP project of HEC Pakistan; the authors also thank Prof. M. S. Kim for reminding them of the scheme in \cite{k-p}.

\bigskip


\begin{thebibliography}{99}
\bibitem{ralph} T. C. Ralph, {\it et al}., Phys. Rev. A {\bf 68}, 042319 (2003).
\bibitem{lund} A. P. Lund, T. C. Ralph, and H. L. Haselgrove, Phys. Rev. Lett. {\bf 100}, 030503 
(2008).
\bibitem{ralph2} T. C. Ralph, Phys. Rev. A {\bf 65}, 042313 (2002).
\bibitem{m-b}W. J. Munro, K. Nemoto, G. J. Milburn, and S. L. Braunstein, \pra {\bf 66}, 023819 
(2002).
\bibitem{gilchrist} A. Gilchrist, {\it et al}., J. Opt. B {\bf 6}, S828 (2004).
\bibitem{g-v} S. Glancy and H. M. Vasconcelos, J. Opt. B {\bf 25}, 712 (2008).
\bibitem{v} D. Vitali, P. Tombesi, and Ph. Grangier, Appl. Phys. B {\bf 64}, 249 (1997).
\bibitem{gerry} C. C. Gerry, \pra {\bf 59}, 4095 (1999).
\bibitem{eit} M. Fleischhauer, A. Imamoglu, and J. P. Marangos. Rev. Mod. Phys. {\bf 77}, 
633 (2005). 
\bibitem{s-i}H. Schmidt and A. Imamoglu, Opt. Lett. {\bf 21}, 1936 (1996).
\bibitem{cc} K. Nemoto and W. J. Munro, Phys. Rev. Lett. {\bf 93}, 250502
(2004); W. J. Munro, K. Nemoto and T. P. Spiller, New J. Phys. {\bf 7}, 137 (2005).
\bibitem{jeong} H. Jeong, \pra {\bf 72}, 034305 (2005).
\bibitem{l-n} S. G. R. Louis, W. J. Munro, T. P. Spiller, and K. Nemoto, \pra {\bf 78}, 022326 (2008).
\bibitem{paris} M. G. Paris, Phys. Lett. A {\bf 217}, 78 (1996).
\bibitem{am} A. P. Lund, H. Jeong, T. C. Ralph, and M. S. Kim, \pra {\bf 70}, 020101(R) (2004).
\bibitem{k-p}M. S. Kim and M. Paternostro, J. Mod. Opt. {\bf 54}, 155 (2007).
\bibitem{andersson} E. Andersson, M. Curty, and I. Jex, Phys. Rev. A {\bf 74}, 022304 (2006).
\bibitem{p-k-h} M. Paternostro, M. S. Kim, and B. S. Ham, Phys. Rev. A {\bf 67}, 023811 (2003).
\bibitem{m-s}W. J. Munro, K. Nemoto, R. G. Beausoleil, and T. P. Spiller, \pra {\bf 71}, 033819 (2005).
\bibitem{chen-yu} Y.-F. Chen, C.-Y. Wang, S.-H. Wang, and I. A. Yu, Phys. Rev. Lett. {\bf 96}, 043603 (2006).


\end{thebibliography}
\end{document}